# Flexural response of all at once 3D printed sandwich composite

Bharath H S[1], Dileep Bonthu[1], Suhasini Gururaja[2], Pavana Prabhakar[3] and Mrityunjay Doddamani[1*]

[1]Advanced Manufacturing Laboratory, Department of Mechanical Engineering, National Institute of Technology Karnataka, Surathkal, 575025, India.
[*]mrdoddamani@nitk.edu.in
[2]Aerospace Engineering, Indian Institute of Science, Bengaluru, 560012, India,
[3]Department of Civil and Environmental Engineering, University of Wisconsin-Madison, Madison, WI, 53706, USA

**Abstract**

Among many lightweight materials used in marine applications, sandwich structures with syntactic foam core are promising because of lower water uptake in foam core amid face-sheets damage. HDPE (high-density polyethylene) filament is used to 3D print sandwich skin, and glass microballoon (GMB) reinforced HDPE syntactic foam filaments are used for the core. The optimized parameters are used to prepare blends of 20, 40, and 60 volume % of GMB in HDPE. These foamed blends are extruded in filament form to be subsequently used in commercially available fused filament fabrication (FFF) based 3D printers. The defect-free syntactic foam core sandwich composites are 3D printed concurrently for characterizing their flexural behavior. The printed HDPE, foam cores, and sandwiches are tested under three-point bending mode. The addition of GMB increases both specific modulus and strength in sandwich composites and is highest for the sandwich having a core with 60 volume % of GMB. The flexural strength, fracture strength, and strain of foam core sandwiches registered superior response than their respective cores. The experimental results are found in good agreement compared with theoretical predictions. Finally, the failure mode of the printed sandwich is also discussed.

## Nomenclature

| Symbol | Description | Symbol | Description |
|---|---|---|---|
| $\rho_c$ | Density - Composite (kg/m$^3$) | $\delta$ | Deflection (mm) |
| $\rho_f$ | Density - Filler (kg/m$^3$) | $I_{eq}$ | Equivalent moment of inertia (mm$^4$) |
| $\rho_m$ | Density - Matrix (kg/m$^3$) | $(AG)_{eq}$ | Shear rigidity (MPa) |
| $V_f$ | Filler volume % | $G_c$ | Shear Modulus (MPa) |
| $V_m$ | Matrix volume % | t | Thickness of Skin (mm) |
| $\phi_v$ | Void content | c | Thickness of Core (mm) |
| $\rho_{th}$ | Density - Theoretical (kg/m$^3$) | d | Distance between centers of skin (mm) |
| $\rho_{exp}$ | Density - Experimental (kg/m$^3$) | $\mu$ | Poisson ratio |
| $E_f$ | Flexural Modulus (MPa) | Y | Distance between base to Centroid axis (mm) |
| L | Span length (mm) | $A_s$ | Cross-sectional Area of Skin (mm$^2$) |
| m | Slope | $Y_s$ | Distance between the neutral axis of the sandwich to Centroid axis of skin (mm) |
| b | Width of the sample (mm) | $A_c$ | Cross-sectional Area of core (mm$^2$) |
| h | Total thickness of the sample (mm) | $Y_c$ | Distance between the neutral axis of the sandwich to Centroid axis of core (mm) |
| $P$ | Load (N) | $I_t$ | Total moment of inertia (mm$^4$) |
| $\sigma_{fm}$ | Flexural stress (MPa) | $I_c$ | Moment of inertia of core (mm$^4$) |
| $E$ | Young's Modulus (MPa) | $I_s$ | Moment of inertia of skin (mm$^4$) |
| $E_s$ | Skin Modulus (MPa) | $\sigma$ | Ultimate Stress (MPa) |
| $E_c$ | Core Modulus (MPa) | $Y_{max}$ | Maximum distance of skin from neutral axis (mm) |
| $V_s$ | Volume % of Skin | M | Moment of resistance (N-mm) |
| $V_c$ | Volume % of core | n | $\frac{E_f}{E_c}$ |

**Introduction**

The development of modern digital manufacturing technology brings a new level of polymer production and offers great flexibility in developing sustainable lightweight construction and multifunctional material systems [1-4]. One of the fastest-growing fields in industrial sectors is Additive manufacturing (AM), which is used for prototyping in initial stages. Today AM has become mainline production for producing many aircraft parts [5, 6], spacecraft components [7, 8], medical devices [9, 10], and consumer products [11]. AM is a layered deposition technique. The most versatile polymer AM technique is 3D printing (3DP), which produces parts by fused filament fabrication (FFF). Because of low-cost printer and the ease of printing parts, 3DP of polymer composite has gained wide commercial success. The most widely utilized polymers in 3D printer are ABS [12, 13], polycarbonate [14], polylactide [15], polymethylmethacrylate [16] and more recently HDPE [17-19]. The enhancement of mechanical properties can be achieved if these thermoplastics are blended with suitable fillers. Some of the commonly used fillers used for blend preparation are Al$_2$O$_3$ [20], glass [21, 22], iron particle [23], fly ash cenospheres [24-27], carbon and glass fiber [28]. The prints realized through these composite blends can exhibit better structural response. Considerable efforts are going on to understand and analyze the quality of the printed parts by varying the different processing parameters. The primary criteria for developing defect-free products in the FFF based 3DP are extruded polymer strands adhesion, solidification of the layer deposited with proper raster diffusion, and part removal post-printing. Nevertheless, dealing with 3DP of semi-crystalline based thermoplastic sandwich composites all at once is challenging due to differential volumetric shrinkage and adhesion issues therein.

The composite printing is an emerging technology to vastly improve the realization of seamlessly integrated components with a substantial reduction in production lead time [29]. Due to its flexibility in design, 3D printing draws the attention of researchers to explore new avenues for composite developments [30-33]. Composite foams (syntactic foams) are produced by incorporating hollow particles in the matrix. Foams are one of those materials known to have lower density and higher damage tolerant morphology [34-36]. Syntactic foams are also referred to as cellular materials, which are categorized into open and closed cell foams. The cells are interlinked in open-cell foams and provide a very higher porosity levels due to struts [36]. The low strength and stiffness in open-cell foams result in low load-bearing capacity of composite foams, and if the skin gets damaged, it results in more moisture absorption in foams. Thereby, closed cell/syntactic foams are utilized prominently in sandwich structures as core [37-41]. Traditional material systems used in automotive, aerospace, civil, and marine structures are replaced by syntactic foams because of its dimensional and thermally stable design coupled with higher load-bearing capacity [42, 43]. The properties of these closed cell foams can be tailored by varying the type of hollow particle, thickness of wall, and volume fraction of filler particles [44-46]. Large numbers of hollow particles are used to manufacture syntactic foam, including glass, carbon, phenol, [47], silicon carbide [48], and alumina. These particles are engineered to get a specified range of diameter and wall thickness for a specific application. Hollow glass particles, commonly referred to as glass micro balloons (GMBs), are extensively used in syntactic foams as fillers and are utilized as the filler in the present investigation. Sandwiches are the special material groups that consist of typically two thin stiffer skins and the lightweight core [49]. The important sandwich characteristics are lower density, higher bending stiffness, damage tolerance, etc. The proper selection of core and skin helps to make sandwiches adaptable to a wider application ranges and different environmental conditions. The selection of skin material is crucial as it comes directly in contact with load and the associated environment. The core structure in sandwich composites are produced by traditional methods like extrusion, expansion, and corrugation limited to geometrically simpler core designs [50]. These conventional techniques do not allow geometrically complex integrated core manufacturing [51-53] necessitating developmental efforts towards 3DP. The sandwich composites manufactured through the conventional approaches as against 3D printed ones have the weakest point across the skin-core interface in addition to limitations of fabricating geometrically complex designed cores. 3DP enabled complex and variable microarchitectures for manufacturing lightweight cellular products [54, 55]. It is necessary to estimate the response of a sandwich in flexure to assess the reliability and safety during its service life. The influence of core topology in bending is investigated theoretically [56, 57]. The flexural response of sandwich with honeycomb cores is elaborately discussed in Ref. [58, 59]. The flexural response of sandwich composites fabricated through conventional manufacturing (core and skin processing individually followed by their assembly through glue) are reported [60-64]. Nonetheless, in authors findings, no investigations are found on bending behavior of 3D printed syntactic foam cored sandwich realized all at once.

In sandwich composites, GMB based foams are most widely used as core due to their higher stiffness and compressive strength [65]. The combo effect of lower density and moisture uptake makes GMB based foams most suitable for realizing floatation modules [66] and submarine buoyancy components [67]. GMB particles with the required size and wall thickness renders greater control over the foams properties. These mechanical properties can be enhanced further if integrated (joint less) syntactic foam core sandwich can be realized all at once. In the present investigation, engineered GMB is used as filler materials as their foams have good physical and mechanical properties compared to fly ash based ones [68, 69]. HDPE is used as a matrix material that finds its applications in milk jugs, chemical containers, household utilitarian, biocompatibility [70], and other structural applications [71, 72]. The composite blends of GMB/HDPE are developed for extruding filaments. HDPE and GMB/HDPE filaments are fed in the printer nozzles to print skin and core respectively for fabricating the syntactic foam core sandwich all at once. The optimized processing parameters are used from Ref. [73]. The 3D printed sandwich composites are investigated for flexural properties. The theoretical prediction is compared with the experimental values, and finally, the failure mode is discussed.

**Materials and Methods**
*Materials and processing*
The matrix HDPE (H) of HD50MA180 grade from IOCL, Mumbai has a melt flow index, Vicat softening point, and density of 20 gm/10 min, 124°C, and 0.950 gm/cm$^3$ respectively. The GMB filler (iM30K) particles are supplied from 3M Corporation, Singapore, having a wall thickness of 1.4 μm and a true density of 0.6 gm/cm$^3$. Both matrix (H) and filler particles are blended at 20, 40 and 60 volume % (H20, H40, and H60) using, Brabender of type 16CME SPL supplied from western company keltron CMEI, Germany, at an optimized screw speed of 10 rpm and blending temperature of 160°C [74]. The GMB volume % is chosen in the range of 20 - 60, as below 20% no appreciable change in mechanical properties is seen, while above 60% volume fraction, much viscous blend formation is noted with particle breakage [75]. Blended GMB/HDPE pallets are processed through a single screw extruder machine of type 25SS/MF/26 supplied by Aasabi machinery Pvt. Ltd., Mumbai having an L/D ratio of 25:1 to develop feedstock filaments of H - H60. The screw-in extruder is set to rotate at a speed of 25 rpm within the heaters with a temperature range of 145-150-155-145°C from the feed to the die segment to melt the polymer blend (H – H60) while moving forward. The semi-viscous mass coming out of extruder is passed through a water tank to be pulled by the take-off unit rotating at 11.5 rpm for manufacturing filaments of 2.85 ±0.05 mm diameter. This parameter minimizes the ovality of the extruded filaments by suitably adjusting the distance between two rollers at the take-off side of the extruder in addition to the speed regulations. Preheating of the blends at 80°C for 24 hours before hopper feeding ensures moisture removal, if any [76]. The extruded H, H20, H40, and H60 feedstock filaments are used in a FFF based printer supplied by Star, AHA 3D Innovations, Jaipur that has 0.5 mm diameter two brass nozzles. The sandwich (S) printing all at once is done by feeding H and H20-H60 filaments in nozzle 1 (N1) and nozzle 2 (N2), respectively, for fabricating SH20 – SH60 syntactic foam core sandwiches. All samples are printed on Kraton$^{TM}$ SEBS FG1901 build plate at bed and chamber temperatures of 120 and 75°C respectively to achieve good adhesion, avoid warpage and to reduce residual thermal stresses. The N1 (225 °C) deposits bottom HDPE skin (1 mm) first. Subsequently, the foamed core is deposited for 6 mm by N2 (225 °C - H20, 245 °C - H40 and H60) [77]. Finally, again N1 prints HDPE skin having a thickness of 1 mm on top of the earlier printed core. G-codes are generated to follow the N1-N2-N1 sequence in order to build sandwich composites (SH20, SH40, and SH60) having a total thickness of 8 mm by using simplify 3D tool path. For all the sandwiches, Y-axis parts orientation, 35 mm/sec printing speed, rectilinear pattern, infill percentage of 100 %, ±45° raster angle, and 0.5 mm layer thickness are set in the printer. The extrusion multiplier that decides deposition volume and is governed by the melt flow index is set at 1 and 1.2 for H - H40 and H60, respectively [78]. Samples are left on the build plate after printing until room temperature is reached. The minimum of five samples of each SH20 - SH60 are printed for characterizing the flexural response.

*Flexural response of sandwich composites*
The foam core sandwiches are printed to the dimensions of 180×18×8 mm$^3$ and are subjected to flexural testing in a three-point bending configuration as schematically presented in Figure 1 (ASTM C393-16). The strain rate of 0.01 s$^{-1}$ (3.41 mm/min crosshead displacement) and 0.1 MPa preload is maintained using Zwick-Roell Z020. Flexural properties are calculated by using,

$$E_f = \frac{L^3 m}{4\, b\, h^3} \quad (1)$$

$$\sigma_{fm} = \frac{3\, P\, L}{2\, b\, h^2} \quad (2)$$

A minimum of five samples are tested, and the average values with standard deviations are reported. Extensive micrography is carried out on as extruded filaments, as printed sandwiches and post-tested prints using a scanning electron microscope (SEM). All the samples are coated with gold sputter covering (JFC-1600) using JSM 6380LA JEOL, Japan. The extruded filaments did not break even after keeping it in liquid nitrogen for 24 hours, and thereby micrographs are taken by cutting them using a knife. The usage of the knife makes the material flow lines clearly visible in the micrographs.

**Results and Discussion**

*Density and void content*

The experimental density of filaments, 3D printed core, and sandwich composites are calculated as per ASTM D792-13 using Contech analytical balance. It is found that in both filaments and printed samples, as the GMB content increases, density decreases. The experimental density of HDPE filament is measured to be 942±8 kg/m$^3$, whereas densities of H20, H40, and H60 foam filaments are noted as 858±15, 780±11, and 683±12 kg/m$^3$ respectively [73]. Similarly, the experimental density of H, H20, H40, and H60, respectively are 927±12, 826±13, 746±18, and 668±10 kg/m$^3$ [73]. It is observed that there is no much variation in densities of filaments and prints, implying GMBs are intact post extrusion and printing processes. The $\rho_{th}$ estimated using the rule of the mixture (ROM) is,

$$\rho_{th} = V_S\rho_S + V_C\rho_C \quad \text{where } (\rho_S = \text{skin density in kg/m}^3) \tag{3}$$

where, $\rho_C = \rho_f V_f + \rho_m V_m$

The ROM results 950, 880, 810, and 740 kg/m$^3$ as $\rho_{th}$ respectively for H, H20, H40, and H60 [73]. The difference between measured experimental density and theoretical densities of both filament and prints results in % void content [73] and is expressed as,

$$\emptyset_v = \frac{\rho_{th} - \rho_{exp}}{\rho_{th}} \tag{4}$$

It is observed that an increase in GMB content increases void %. The voids in filaments and printed cores of H-H60 ranges between 0.84-7.70 and 2.42-9.73%, respectively. The rise in void content post-printing indicates that the filament porosity is retained in prints and possibly might have elongated during the printing process, making the syntactic foams a three-phase structure (HDPE, GMB, voids). Such three-phase syntactic foams might enhance energy absorbing capabilities further. The weight saving potential of printed foams is estimated to be in the range of 10.9-27.94% as compared to H [73]. As seen from Table 1, density of sandwich decreases as GMB content increases. SH20-SH60 densities are higher (6.45-8.36%) compared to H20-H60 counterparts and are expected due to the additional HDPE skin on the foam cores. The maximum weight saving potential is noted to be ~22% in SH60, implying the developed syntactic foam core sandwich using 3D printing can replace a few of the components in buoyancy modules having enhanced specific mechanical properties with integrated (without any joint) complex geometrical features.

*Microstructural characterization*

The scanning electron microscopic image of the GMB particle, which is used as filler in foams, is shown in Figure 2a. The surface characteristic of the GMB particle is captured at higher magnification, which shows a very smooth surface without any defects. Figure 2b presents intact GMBs post blending for representative H60 composition. The micrograph of the knife cut representative H20-H60 filaments (Figure 3) shows intact uniformly dispersed GMB particles with poor interfacial bonding. The printed syntactic foam core sandwiches are freeze fractured, and the micrographs are shown in Figure 4. The seamless bonding at the skin-core interface in all the representative printed sandwiches is clearly visible from these micrographs, implying the suitability of the printing parameters utilized in the present work. With the chosen printing parameters, SH20-SH60 sandwiches are printed, and a representative micrograph of SH60 print across three different zones from top to bottom skin is presented in Figure 5a. Figure 5b presents the micrograph in the thickness direction. Both these micrographs clearly indicate seamlessly diffused layers across and along with the prints. The printed SH60 image is presented in Figure 5c show the successful demonstration of sandwich printing all at once, completely eliminating adhesive joining of skin and cores like in conventional manufacturing.

*Flexural response of 3D printed core and sandwiches*
The flexural test is carried out in a three-point bending configuration where core and sandwich samples are mounted, as shown in Figure 6a. With the gradual application of load, the sample starts to yield, as shown in Figure 6b. Among foams, H40 and H60 showed brittle fracture as compared to H and H20, which did not fail until 10 % strain. Brittleness is due to the inclusion of GMB in HDPE. In H40 and H60 foams, crack initiated from the tensile side propagated along the direction of loading until it met the compressive side (Figure 6c), indicating perfectly diffused and bonded layers. An increase in GMB content increases modulus as shown in Figure 8a. The modulus of H, H20, H40, and H60 is 990±11.28, 1210±19.56, 1280±11.87, and 1360±11.23 MPa, respectively. It is observed that the modulus of H60 is 1.37 times higher than H, which is due to intact GMB particles even at the highest filler loading (Figure 7c). The extensive plastic deformation is seen at lower filler contents (Figure 7a and Figure 7b). The strength of H, H20, H40, and H60 are found to be 25.4±0.12, 21.0±0.58, 17.1±0.47, and 15.1±0.72 MPa respectively [73], where HDPE has the highest strength compared to foam samples, which is 1.20, 1.48, and 1.68 times higher than H20, H40, and H60 foams strength. With increasing GMB content, the flexural strength of core decreases, as seen from Figure 8b. Nonetheless, specific properties need careful attention along with flexural response evaluation of such foam cored sandwich constructions. The filler inclusion increases amorphous fraction resulting in more restrained matrix flow and mobility of the polymer chain leading to weaker interfaces. Improving the bonding between the constituents through appropriate coupling agents may increase the strength but at the expense of a significant reduction in ductility that may obstruct filament extrusion and 3D printing process.

In the bending test of sandwich samples, the stress varies across the sample thickness from compression (top skin where the loading wedge touches the specimen) to tensile (bottom skin) side. Additionally, the shearing stress act along the specimen's length predominantly in conventionally manufactured sandwich composites leading to skin-core debonding and subsequent failure. Therefore, locations of crack origin and directions of propagation helps in determining the types of stresses causing failure. Figure 9 presents yielding and maximum mid-point deflection of representative SH20. SH20 did not fail until 10 % strain and, as anticipated, registered the highest strength as compared to other sandwiches. SH40 and SH60 showed a brittle fracture (Figure 10a). In sandwich composites, the crack is initiated in the bottom HDPE skin and later propagated along with the core just below the loading point. Failure begins at the specimen's tensile side, just below the loading point, and develops toward the compressive side. For all the 3D printed syntactic foam core sandwiches, slimier failure features are observed owing to the suitable printing parameters used, avoiding shear crack/failure along with the printed layers. The modulus increases with GMB content (Table 2 and Figure 10b). SH60 showed the highest modulus compared to other sandwich compositions. Intact GMBs at higher filler loading, as clearly evident from Figure 11b enhances the moduli of SH60. With increasing GMB content in the core, flexural strength decreases, as seen from Figure 10. SH20 and SH40 failed completely in two pieces exhibiting the typical brittle fracture. SH20 is the best in strength, which might be due to effective load transfer between the constituents. This observation is based on the absence of plastic deformation of HDPE, as seen in Figure 11a. The excessive plastic deformation of the matrix at higher filler loading makes SH60 perform lower as compared to SH20. Nonetheless, the specific strength of SH60 is 1.1 times higher than that of SH20. In the case of SH40 and SH60, the crack has initiated near the mid-span of the specimen and propagated vertically across the thickness of the core and reaches the upper HDPE skin. The top skin's progressive failure reduces the rate of drop in stress and provides extra strain before failure. The interfacial failure is a common thing in shear stress that influenced sandwich composites [73]. Nevertheless, in the one-shot printed syntactic foam core sandwiches as presented in this paper, none of them exhibited interfacial separation between core and skin owing to perfect and seamless bonding (Figure 5). Figure 12 presents a flexural strength comparison between printed core and respective sandwiches. Though strength is seen to be decreasing with GMBs addition, specific flexural strength increases and is a crucial factor in weight-sensitive structural applications. The flexural strength of SH20, SH40, and SH60 is 1.05, 1.22, 1.35 times higher than their respective H20, H40, and H60 cores, indicating the potential benefit of realizing all at once 3D printed syntactic foam core sandwich. Based on the experimental investigations in this study, SH60 has the highest specific modulus and strength values, which can be exploited for potential weight applications without compromising the mechanical properties.

*Theoretical prediction of sandwich properties*

Theoretical values of modulus and failure load of the printed syntactic foam core sandwiches are estimated using properties of the skin and core evaluated individually using an experimental approach. The terminologies used for theoretical predictions and comparative load-deflection plots for printed sandwiches are presented in Figure 13. In flexure loading condition, the load is applied gradually at the beam center and, the deflection includes deformation of both the skin and core. The mechanical properties of top skin degrade in multi-axial stress presence when the wedge comes directly in contact with top skin. Therefore the thickness of top skin where the load is applied is neglected in theoretical calculations of deflection [79, 80]. This deflection can be calculated using Eqn. 5 [81]. The effectiveness of skin-core bonding on the properties of the 3D printed sandwich structure can be estimated by the theoretical modulus of the sandwich which is calculated using the ROM (Eqn. 9).

$$\delta = \frac{PL^3}{48(EI)_{eq}} + \frac{PL}{4(AG)_{eq}} \tag{5}$$

Here $EI_{eq}$ is called flexural rigidity and $(AG)_{eq}$ is shear rigidity.

$$EI_{eq} = \frac{bt^3 E_s}{12} + \frac{btd^2 E_s}{4} + \frac{bc^3 E_c}{12} \tag{6}$$

$$AG_{eq} = \frac{bd^2 G_c}{C} \tag{7}$$

$$G_c = \frac{E_c}{2(1+\mu)} \tag{8}$$

$$E = (E_s V_s) + (E_c V_c) \tag{9}$$

The skin and core moduli are 730±20.54 MPa and experimental results extracted from Ref. [73], respectively. Table 2 lists experimental and theoretical flexural modulus values and is observed to be in good agreement when compared. The deviations between the theoretical and experimental results (9.99-11.45%) are attributed to void contents in the printed sandwiches. The failure load evaluation of the sandwich structure depends on the neutral axis (Eqn. 10) and the total moment of inertia (Eqn. 11). As the loading condition in the three-point bending test is simply supported, the moment at the center is considered, and by using Eqn. 13, the critical load is evaluated.

$$Y = \frac{(A_s E_s Y_s) + (A_c E_c Y_c)}{(A_s E_s) + (A_c E_c)} \tag{10}$$

$$I_t = \frac{(E_c I_C + E_s I_s)}{E_c} \tag{11}$$

$$\sigma_{fmax} = \frac{nMY_{max}}{I_t} \tag{12}$$

$$M = \frac{P}{2} \times \frac{L}{2} \tag{13}$$

Table 3 presents the theoretical and experimental critical load estimations for printed sandwiches and is noted to be decreasing with increasing GMB content, which might be due to again higher void content at higher filler loadings. The deviation between the experimental and theoretical loads is noted to be in very close good agreement up to half of the maximum load (Figure 13b). Such theoretical approaches help in predicting the sandwich properties beforehand, which in turn decides a broad range of possible applications. The load-deflection cure for experimental and theoretical predictions is represented in Figure 13b.

*Failure mode of sandwich structure*
The type of sandwich failure strongly depends on skin geometry, strength and core material [49, 82]. The three possible failure modes in sandwich composites under flexure are indentation, shear, and micro buckling/face wrinkling. The sandwich faceplates remain elastic during core indentation and shear failures [83]. Indentation creeps in when compressive yield strength matches with stresses developed through the thickness of the core, as shown in Figure 14a. The plastic indentation zone (λp - core reactive force equals core compressive strength) and elastic indentation zone (λe - reactive force equals *kw*) forms the total indentation region. In case of shear failure, radial shear strain in core exceeds failure strain. In earlier efforts faceplates contribution is ignored [84, 85] while circumferential hinges work is accounted for their consideration [86]. The bottom skin fails first as it is subjected to tension while micro buckling/face wrinkling surfaces on the top skin (compressive side). Sandwich structures with ductile skins failed in the bottom skin, while those with the brittle ones failed with micro-buckling in the top skin [87]. Generally, when the load is applied to the sandwich structure, the skin undergoes tensile/compressive failure, whereas core undergoes shear failure. The shear is not observed for all the tested sandwiches. A linear indentation is observed at the point where wedge directly comes in contact with top skin and when the load gradually increases, compressive stresses are induced on the top skin resulting in wrinkling at the center for SH20 (Figure 14c). In the present work the indentation failure is observed in SH40 and SH60 samples, as seen from the representative image in Figure 14d. None of the samples failed in shear. All the samples except SH20 fractured just below the loading point in an approximately straight line, as seen from Figure 14d. The indention is located in the marked area of Figure 14d on the fractured surface of the top skin. Also, crack initiation at the bottom skin, and shear failure of the core is observed in SH40 and SH60 due to a higher amount of stiffer GMBs inclusion leading to brittle behavior. Similar failure features except shear are observed for the printed sandwiches developed in the present work [87-91].

**Conclusion**
The present work dealt with the flexural response of all at once 3D printed GMB/HDPE syntactic foam core sandwich. The results are summarized as follows:
- The suitable printing parameters employed for all at once 3D printed syntactic foam core sandwich resulted in seamless bonding at the skin-core interface.

- The printed SH60 has a weight-saving potential of ~22%.
- Voids in the core enhance energy absorbing capabilities and make them three-phase syntactic foams.
- SH60 sandwich exhibits the highest specific modulus and strength.
- 3D printed sandwich has superior strength and is in the range of 1.05-1.35 times as compared to their respective foam cores.
- Shear failure, which is very common, is not observed in 3D printed sandwich constructions.
- Experimental results are in good agreement with theoretical predictions.

The higher specific mechanical properties of 3D printed syntactic foam core sandwiches compared to core counterparts as observed in the present work, opens new avenues of exploring different skin and core combinations in addition to improving the interfacial bonding between the constituents by suitable filler surface treatments. Further, 3D printing makes the possibility of realizing joint less (leak proof) geometrically complex sandwich constructions, which shall act as a great boon in marine, automobile, and aerospace sectors.

**Acknowledgment**
The authors wish to acknowledge the support of the SPARC grant (SPARC/2018-2019/P439/SL), Govt. of India. The authors would also like to acknowledge the Mechanical Engineering Department of the National Institute of Technology, Karnataka, Surathkal, for providing the facilities and support for conducting research work. The authors also wish to acknowledge U.S. Office of Naval Research - Young Investigator Program award [Grant No.: N00014-19-1-2206] for support towards the work presented here.

**Data availability**
The raw/processed data required to reproduce these findings cannot be shared at this time as the data also forms part of an ongoing study.

Table 1. Density, void % and weight saving potential estimations of printed sandwich.

| Material | $V_f$ | $\rho_{exp}$ | $\rho_{th}$ | $\phi_v$ | Weight Saving Potential (%) w.r.t H |
|---|---|---|---|---|---|
| SH20 | 20 | 879.35±14 | 897.5 | 2.02 | 5.14 |
| SH40 | 40 | 777.38±16 | 845 | 8.00 | 16.14 |
| SH60 | 60 | 723.87±11 | 792.5 | 8.66 | 21.91 |

Table 2. Flexural response of prints.

| Materials | Experimental Modulus (MPa) | Theoretical Modulus (MPa) | Strength (MPa) | Fracture strength (MPa) | Fracture strain (%) |
|---|---|---|---|---|---|
| SH20 | 927±18.46 | 1067.83 | 21.80±0.45 | ----- | ----- |
| SH40 | 1000±13.58 | 1126.09 | 20.53±0.52 | 20.25±0.57 | 7.13±0.15 |
| SH60 | 1050±12.86 | 1186.57 | 19.72±0.80 | 19.72±0.77 | 5.20±0.10 |

Table 3. Experimental and theoretical critical load estimations.

| Material | Experimental Critical load (N) | Theoretical Critical load (N) from Eqn. 13 | Deviation (%) |
|---|---|---|---|
| SH20 | 135 | 138.67 | 2.57 |
| SH40 | 133 | 138.57 | 4.10 |
| SH60 | 118 | 135.60 | 12.97 |

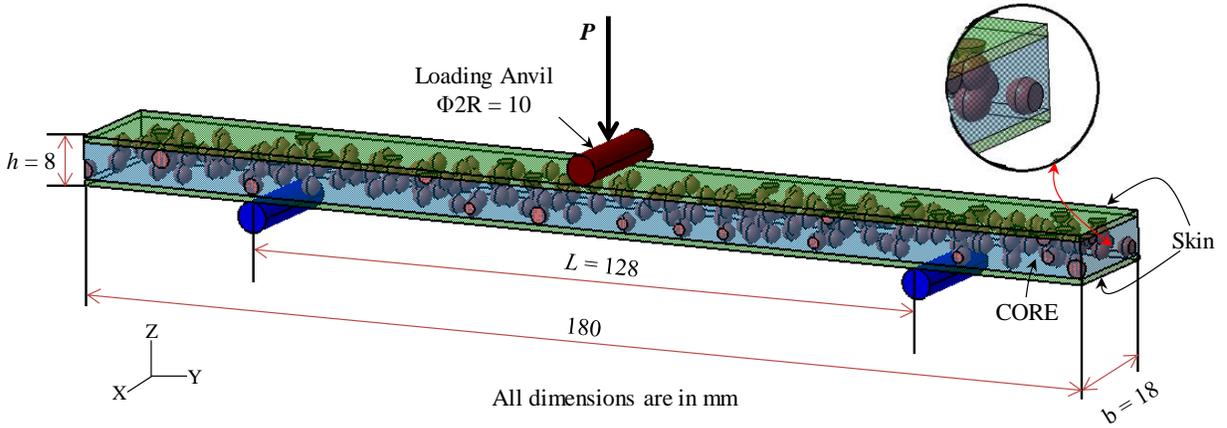

Figure 1. Print dimensions and flexural test configuration.

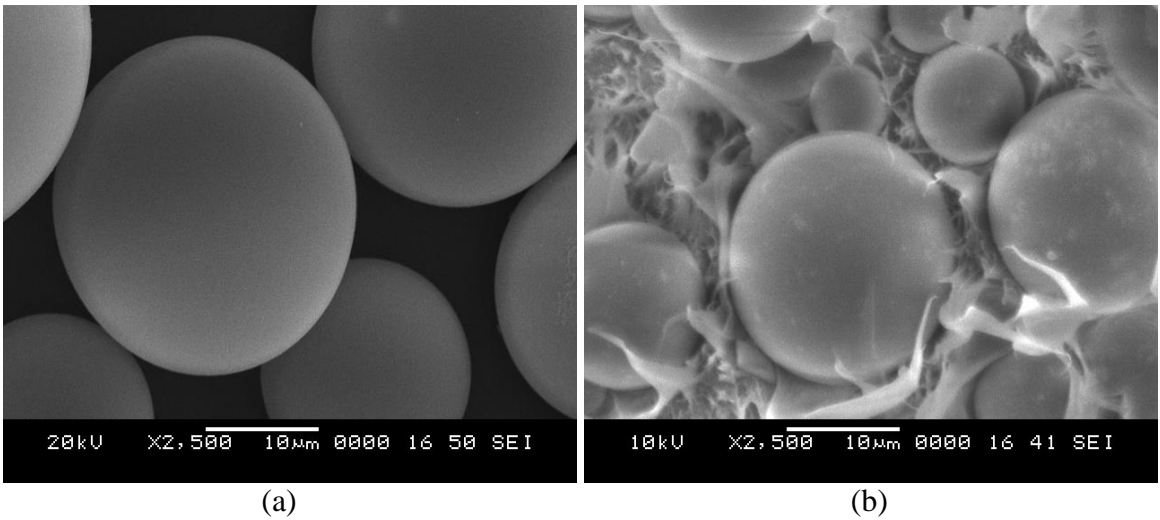

(a)                                                             (b)
Figure 2. Micrograph of (a) GMB particles and (b) GMB/HDPE blend (H60).

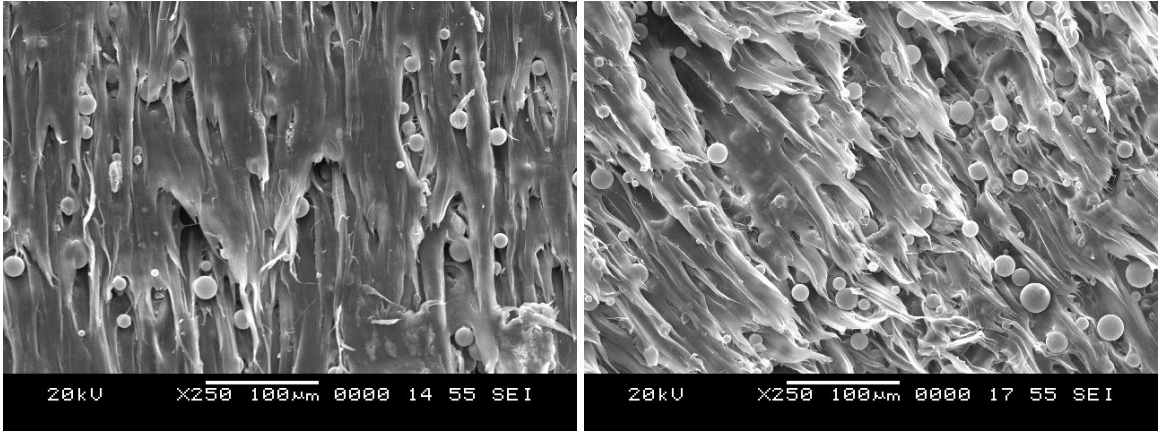

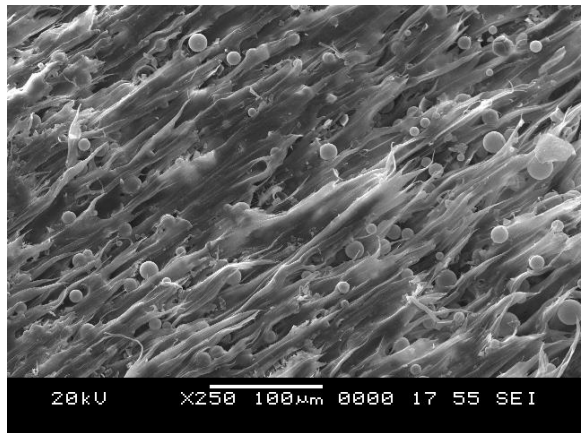

Figure 3. Cross-sectional SEM micrographs of (a) H20, (b) H40 and (c) H60 filaments.

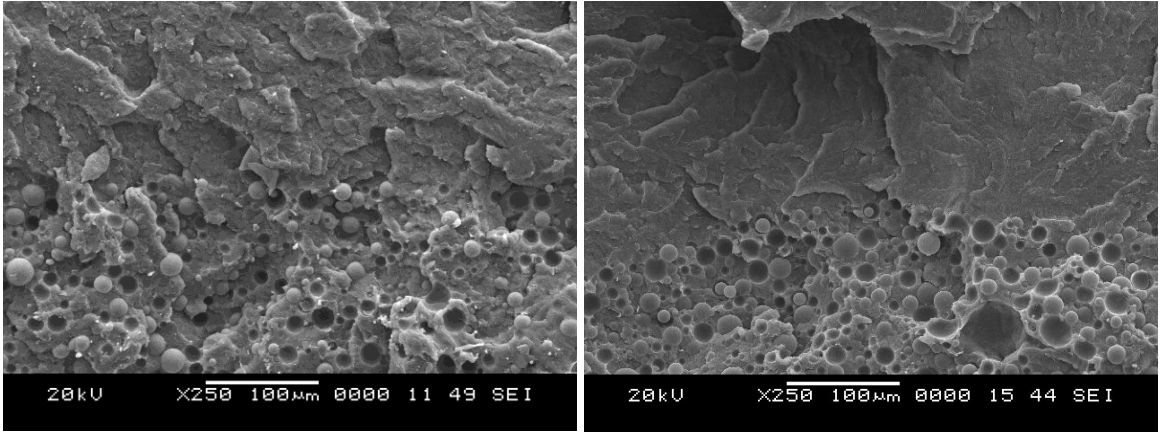

(a) (b)

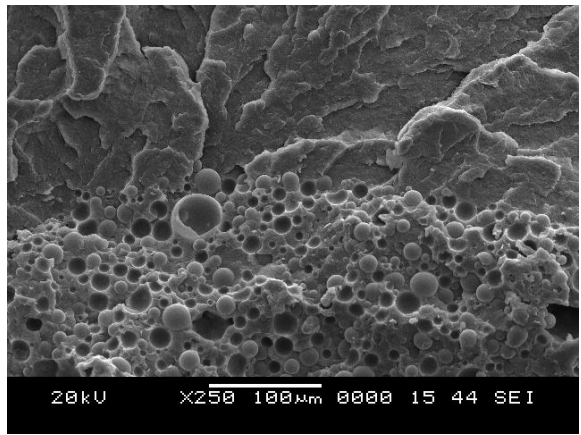

(c)

Figure 4. Freeze fractured micrographs of (a) SH20, (b) SH40 and (c) SH60 at skin-core interface.

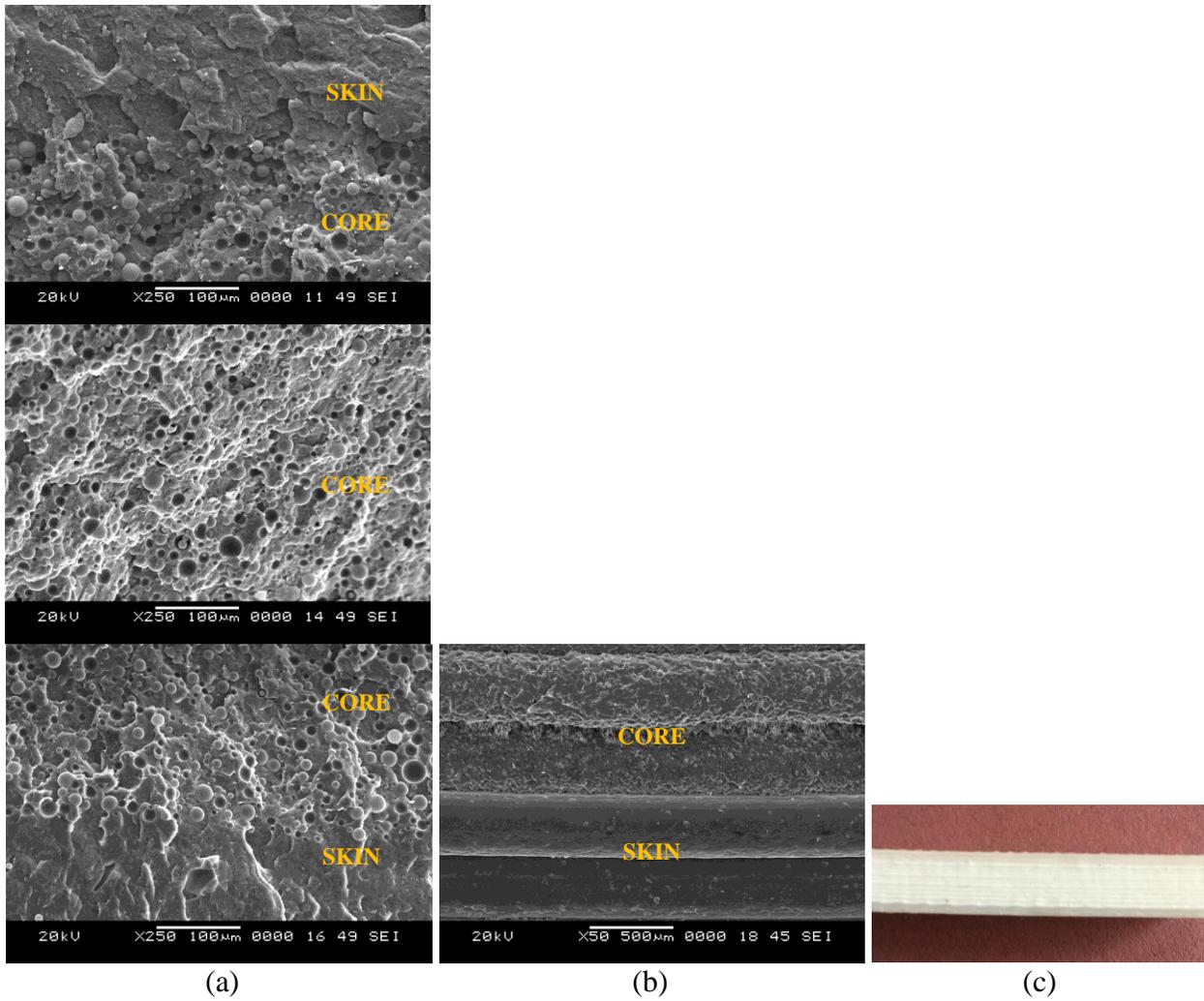

(a)                                     (b)                                     (c)

Figure 5. As printed freeze fractured micrograph of sandwich (a) across the thickness (b) along the thickness. (c) representative printed SH60 sandwich composite.

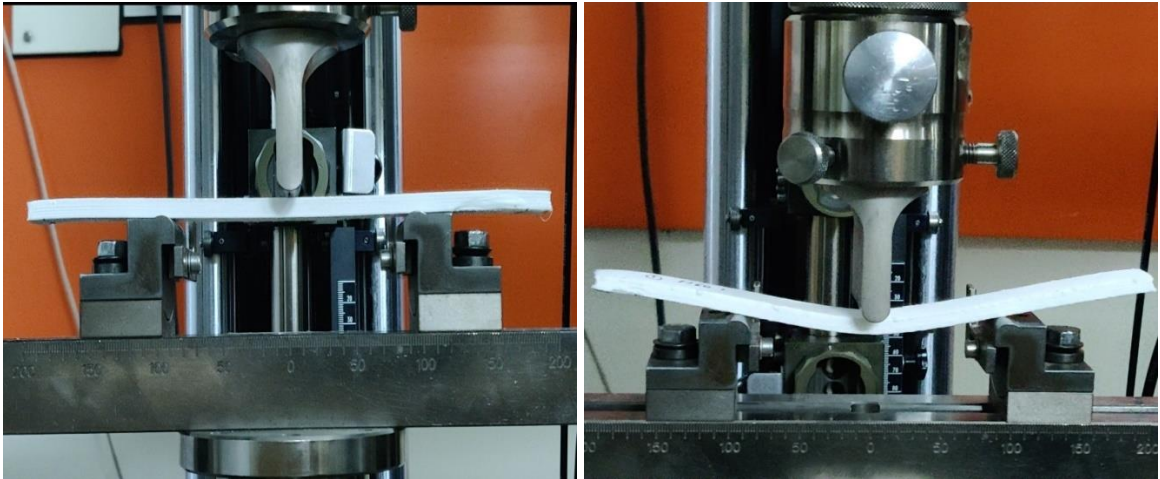

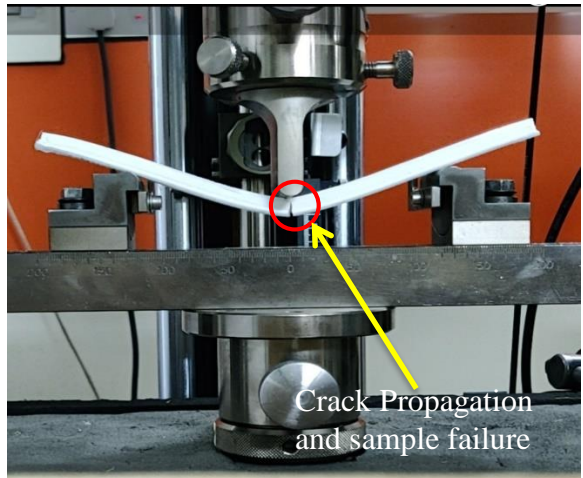

Figure 6. (a) Representative H60 mounting in flexure mode (b) yielding (c) and crack initiation.

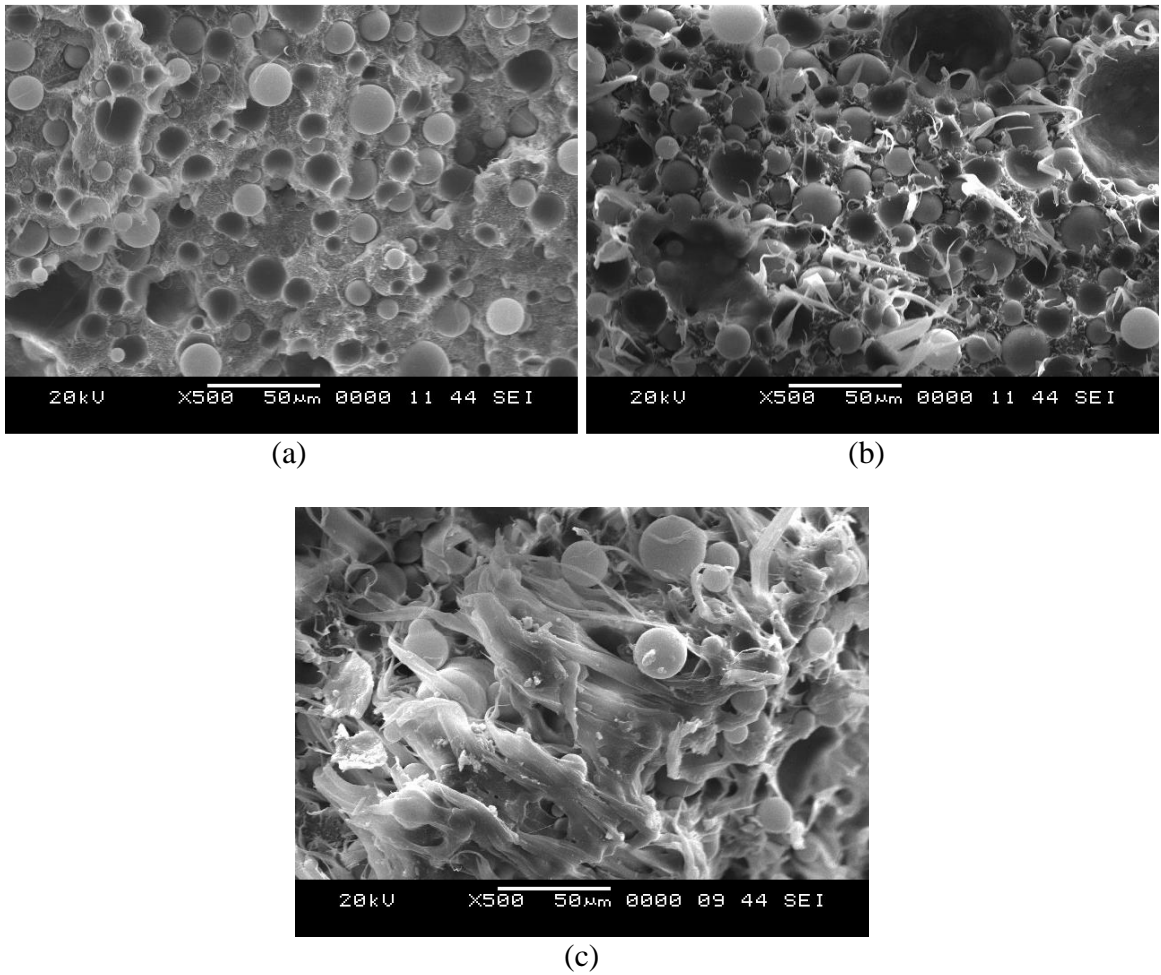

(a)                  (b)

(c)

Figure 7. Micrographs of post flexure tested (a) H20 (b) H40, and (c) H60 printed cores.

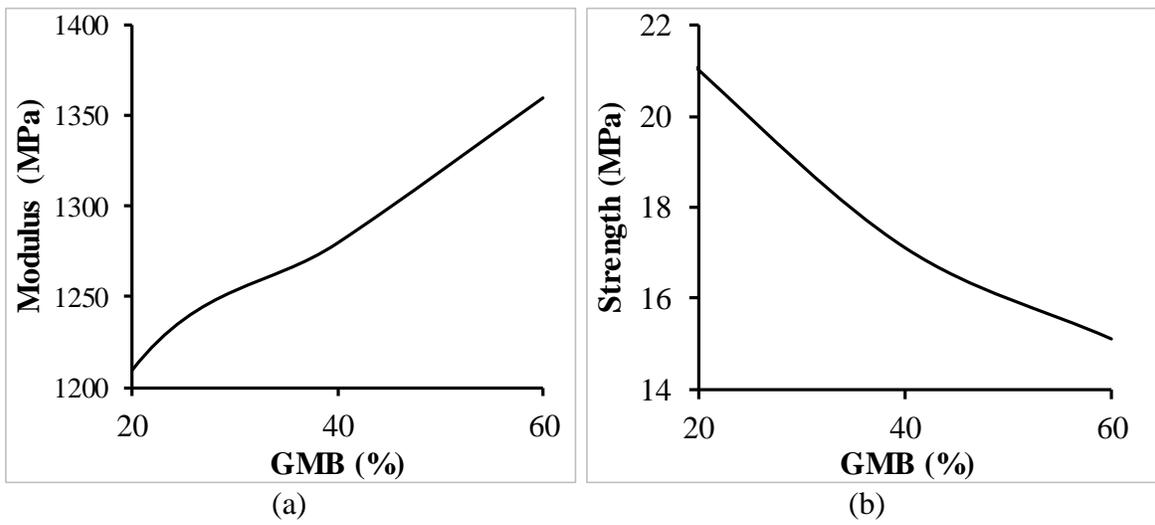

(a)                  (b)

Figure 8. (a) Flexural Modulus and (c) strength as a function of GMB content for H20-H60.

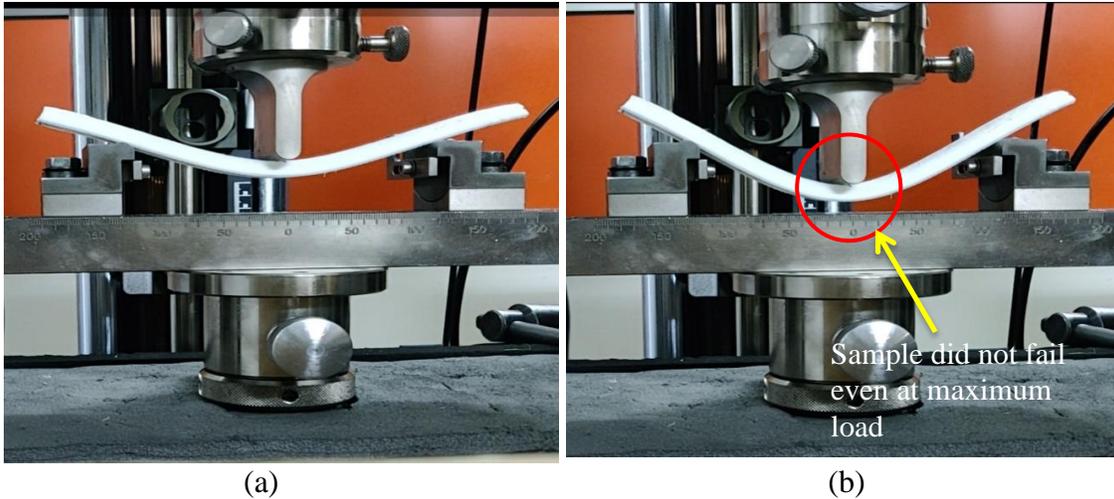

(a)                                (b)

Figure 9. Flexural test of representative SH20 (a) yielding and (b) maximum mid-point deflection.

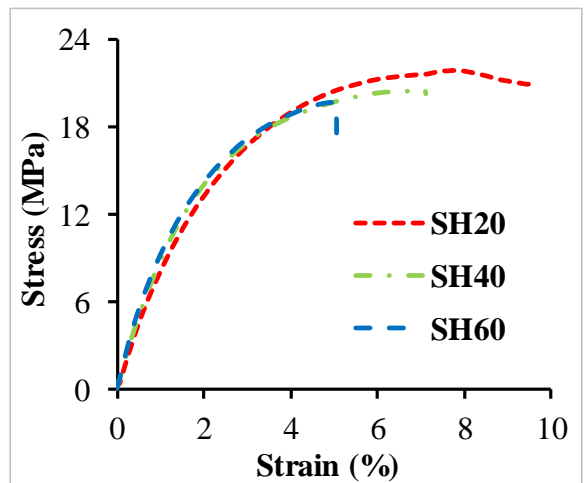

(a)

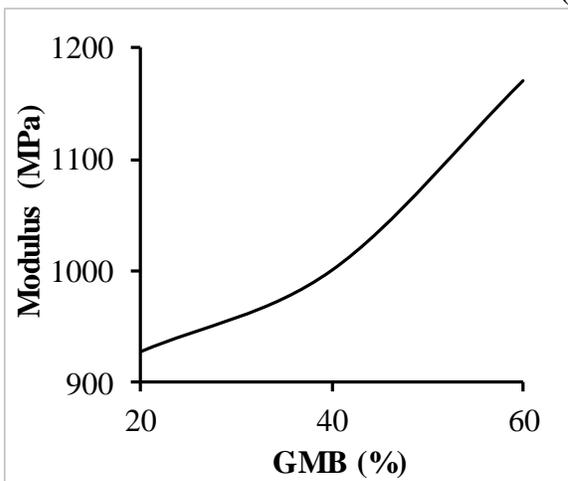 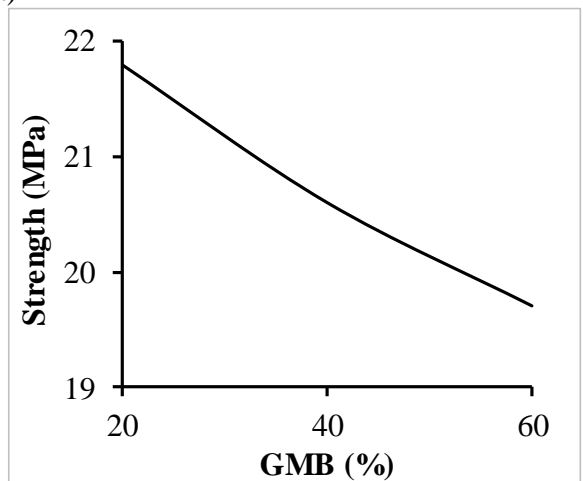

(b)                                (c)

Figure 10. (a) Stress - strain plots (b) Modulus and (c) strength as function of GMB content in printed sandwiches.

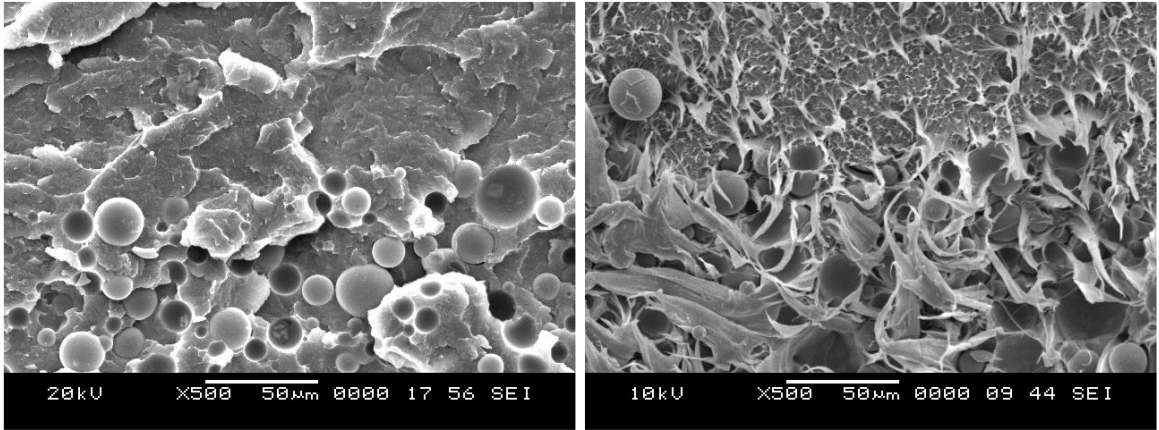
(a)                                         (b)

Figure 11. SEM of post flexure tested representative printed (a) SH20 and (b) SH60.

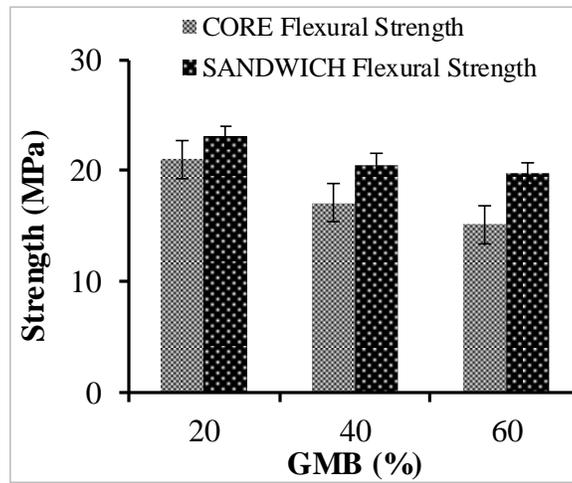

Figure 12. Printed Core and Sandwich comparison for strength.

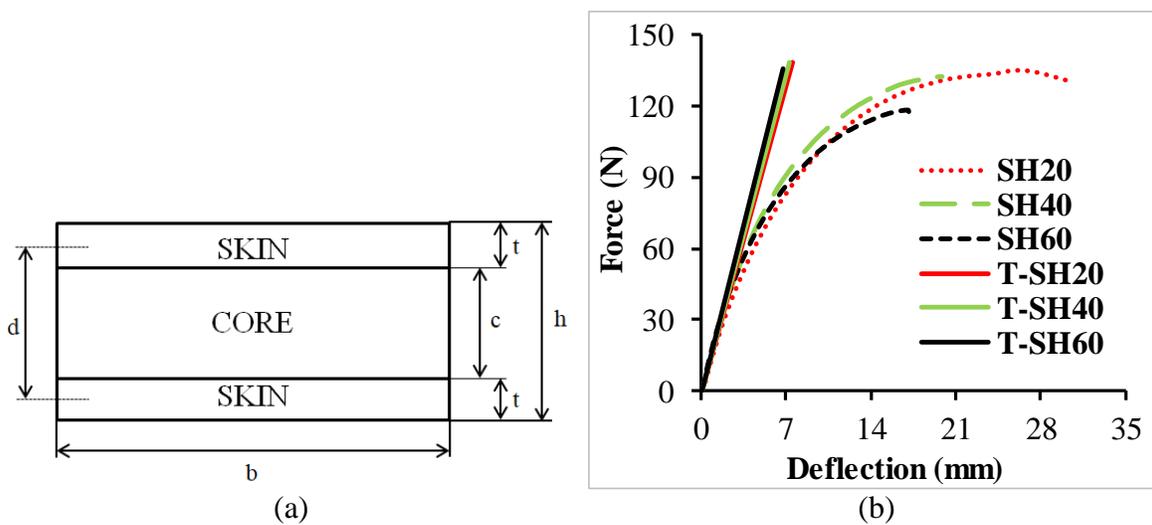

(a)                                         (b)

Figure 13. (a) Schematic representation of sandwich with the terminologies used and (b) comparative force-deflection plots for theoretical and experimental approaches. T denotes "theoretical".

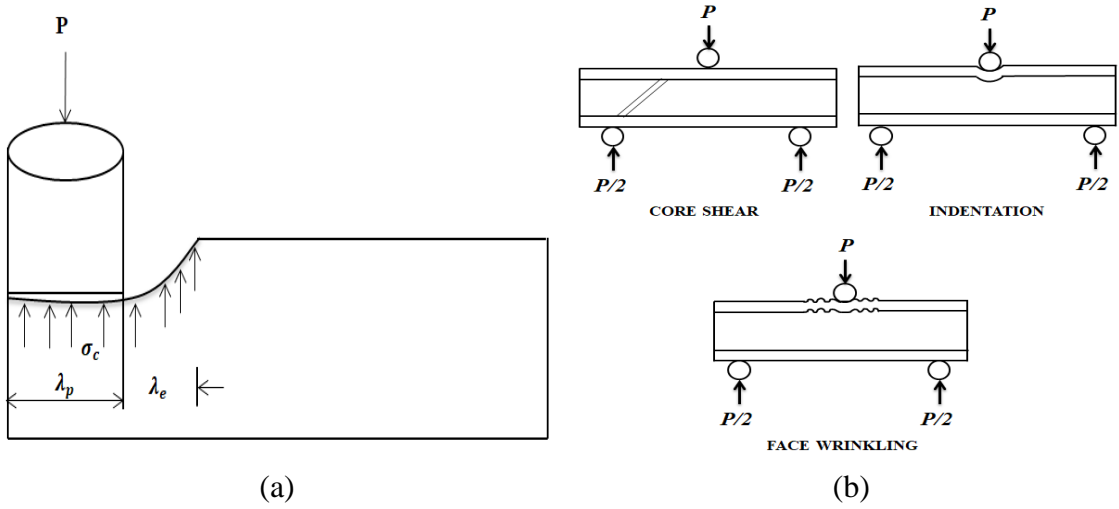
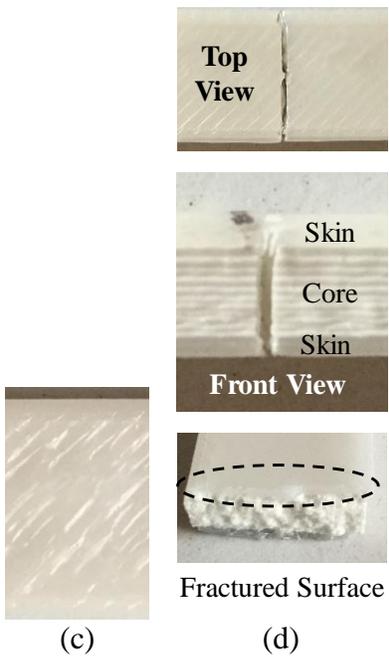

Figure 14. Schematic representation of (a) core indentation and (b) failure modes observed in 3D printed syntactic foam core sandwiches. (c) face wrinkling in SH20 and (d) indentation failure (SH40 and SH60).